\documentclass[12pt,a4paper]{article}
\usepackage{amsmath,amssymb}
\usepackage{epsfig}
\begin{document}
\begin{center}
\Large\textbf{ Comments on astro-ph/0305039 and gr-qc/0309036}\\
\bigskip
\normalsize A. V. Minkevich\\
\medskip
 \textit{\small Department of Theoretical Physics, Belarussian State University,
av. F. Skorina, 4, 220050, Minsk, Belarus }
\end{center}
\begin{center}
\begin{minipage}{0.75\textwidth}\small
\textbf{Abstract.}The main results of  papers gr-qc/0307026 and gr-qc/0312068 are
formulated. These results are opposite to conclusions of paper astro-ph/0305039 and
comments gr-qc/0309036.
\end{minipage}
\end{center}

Our papers \cite{com1,com2} are devoted to study inflationary cosmological models by
using cosmological equations for homogeneous isotropic models of gauge theories of
gravitation - so-called generalized cosmological Friedmann equations (GCFE). The
investigation of homogeneous isotropic models filled by usual gravitating matter
carried out in a number of our previous papers (references are given in
\cite{com1,com2}) showed that their generic feature is regular bouncing character,
if equation of state of gravitating matter at extreme conditions (extremely high
energy densities, pressures) and indefinite parameter $\beta$ of GCFE satisfy
certain restrictions. By including scalar fields mathematical structure of GCFE is
essentially changed, namely, particular solution of GCFE determined by equation
$Z=1-\beta(\rho-3p)=0$  ($\rho$ is energy density, $p$ is pressure) becomes
nonstationary, and the bounce point (corresponding to $Z=0$ in the case without
scalar fields) is transformed into families of bounce curves separated from
particular solution. The analysis of GCFE (with $\beta<0$) by including scalar
fields leads us to the following conclusions:
\begin{enumerate}
\item  The greatest part of inflationary cosmological solutions including scalar
fields and ultrarelativistic matter are regular in metrics and Hubble parameter and
have bouncing character \cite{com1}; singular solutions appear because of divergence
of particular solution, however, their number is essentially less than the number of
regular solutions, namely, infinite number of regular solutions correspond to each
singular solution.
\item  The problem of excluding noted above singular solutions
was analyzed and solved in Ref \cite{com2}. It is shown that all inflationary
cosmological models including also usual gravitating matter with equation of state
$p_1=p_1(\rho_1)$ besides scalar fields are regular in metrics and Hubble parameter
and have bouncing character, if gravitating matter at extreme conditions  satisfies
the following restriction $\frac{1}{3}\,\rho_1<p_1\le\rho_1$. In this case regular
inflationary cosmology can be built within the framework of gauge theories of
gravitation.
\end{enumerate}
Noted results were obtained for flat, open and closed models in the case of
various scalar field potentials applying in chaotic inflation. In particular
case of flat models by using scalar field potential $V=\frac{1}{2}\,m^2\phi^2$
discussed in Refs \cite{com3,com4} our results have to be valid also.

\end{document}